\documentclass[galaxies,article,referee,oneauthor,pdftex,12pt,a4paper]{mdpi}
\usepackage{graphicx}
\RequirePackage{color}
\makeatletter
\renewcommand\@biblabel[1]{#1. }
\makeatother
\def\aap{A\&A\,  }

\def\araa{ARA\&A  }

\def\jcap{Journal of Cosmology and Astroparticle Physic  } 
 
\def\mnras{MNRAS\,  }

\def\prd{Phys. Rev. D   }
\def\2F1{~_2F_1}
\lastpage{x}
\doinum{10.3390/------}
\pubvolume{xx}
\pubyear{2015}
\history{Received: xx / Accepted: xx / Published: xx}
\Title
{
Classical and relativistic evolution of an extra-galactic 
jet with back-reaction
}

\Author{Lorenzo Zaninetti $^{1,}$}

\address
{
$^{1}$
Department   of Physics, 
University of Turin,
via P.Giuria 1,\\ I-10125 Turin,Italy
}

\corres
{
zaninetti@ph.unito.it
}

\abstract
{
We consider a turbulent  jet which is moving in a 
Lane--Emden ($n=5$) medium.
The conserved  quantity is  the  energy    flux,
which allows finding,  to first order, 
an analytical expression for the 
velocity and an approximate trajectory.
The conservation of the relativistic 
flux for the   energy
allows deriving, to first order,  an analytical 
expression for the velocity, 
and 
numerically determining the trajectory.
The back-reaction due to the radiative losses for the trajectory 
is evaluated  both in the classical and the relativistic case.
}



\begin{document}

\section{Introduction}

\begin{figure*}
\begin{center}
\includegraphics[width=7cm]{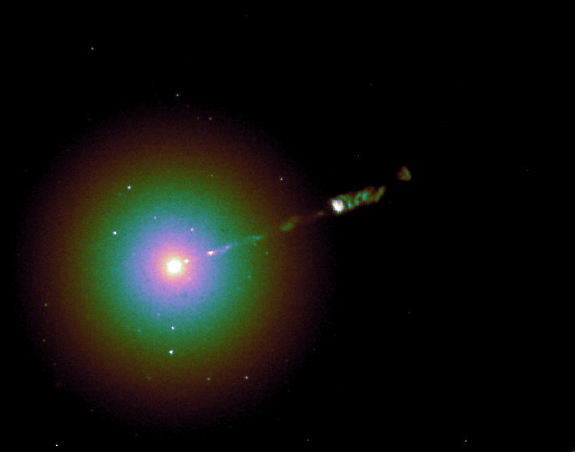}
\end {center}
\caption
{
The super-giant elliptical galaxy M87 and the optical jet; 
the credit is due to  Instituto de Astrof\'{i}sica de Canarias. 
}
\label{m87}
    \end{figure*}

The study of extra-galactic jets started with the observations
of NGC 4486 (M87), where
`a curious straight ray lies in a sharp
gap in the nebulosity \ldots',
see \cite{Curtis1918} and Figure~\ref{m87}.
At the moment of writing, 
the extra-galactic radio sources are classified   on the basis of
the position  of the brightest radio emitting regions with respect
to the channel, see~\cite{fanaroff,Kembhavi1999} for details.
FR-I,  after Fanaroff \& Riley,
 have hot spots that are more distant from the
nucleus (a typical example is Cygnus~A) 
and  luminosity, $L$, at 178 Mhz of 
\begin{equation}
L \leq h_{100}\, 2\,10^{25} \frac{W}{Hz\,str}  \quad \text{FR-I}
\quad,
\end{equation}
where $h_{100}=H_0/100$ and $H_0$  is the Hubble constant.
FR-II radio galaxies have
emission uniformly  distributed along the channel (typical example
3C449)
and  luminosities greater than the above value
or, in other words, the more powerful radio galaxies are 
classified as FR-II.
A list  of the properties, length in kpc and  power  in Watt, 
of  extra-galactic radio jets  can be found in
\cite{Bridle1984,Liu2002}.
In the following we will study jets with small openings, 
such as that of M87.

The  problem  of the velocity  of extra-galactic radio jets  
has  been  analysed in two ways:
\begin{enumerate}
\item
The velocity  of the jet  is constant  over  many kpc
and takes the value  $v$.
Due to the fact that is thought that this  velocity 
is nearly  relativistic, it  is parametrized
as $\beta=\frac{v}{c}$, where $c$ is the velocity of light.
As an example, \cite{Hardcastle2004} analysed  some
wide-angle tail radio galaxies 
and found a terminal velocity of $\beta=0.3$.
\item  
The velocity of the jet decreases  with  an imposed law,
see  \cite{laing2002}, 
or  is evaluated by a numerical code,
see  \cite{Nawaz2014,Nawaz2016}.
In this case, the relativistic parameter $\beta$ decreases 
along the trajectory.
\end{enumerate}
Recently the problem of the decrease of the velocity along  a turbulent
jet has been solved, imposing the conservation of the flux of momentum,
see \cite{Zaninetti2015d}, or
imposing the conservation of the   energy    flux, see 
\cite{Zaninetti2016e}.
The approach using the conservation of the   flux
of energy is attractive because it has the same dimension
of the luminosity.
Further on, the jets are radiating away 
in the various  observational bands, such as radio, optical, infrared,
etc., and 
we briefly recall that the extra-galactic radio 
source covers a range in observed luminosity
from $10^{19} \frac{W}{Hz}$  to $10^{28} \frac{W}{Hz}$,
see \cite{Kellermann1998}.

Therefore the   flux of energy available at the beginning
of the jet 
will progressively decrease due to the radiative losses.
This  paper, in Section \ref{secclassic},
introduces the Lane--Emden  ($n=5$) density profile
and consequently derives  an
approximate trajectory to first order 
as well  as a numerical trajectory to second order in the presence of losses.
In Section \ref{secrelativistic}
we present 
a series solution for the relativistic trajectory to 
first order and 
a numerical solution to second order.
Section \ref{secapplication} models 
the intensity of the radio-jet  in 3C31.

\section{Conservation of the flux of energy }
\label{secclassic}

A turbulent jet is defined as a jet  
which has the same density 
as the surrounding intergalactic medium (IGM), see the next
subsection for details.
The conservation of the   flux of energy   in a turbulent jet
has been explored in \cite{Zaninetti2016e}
for three types of IGM,  with the following
radial dependences:
constant density profile, 
hyperbolic and inverse power law density profiles.
Here we analyse the case of a Lane--Emden (n=5) density
profile, to which a subsection will be dedicated.

\subsection{The turbulent jet}

\label{turbjet}
\begin{figure*}
\begin{center}
\includegraphics[width=7cm]{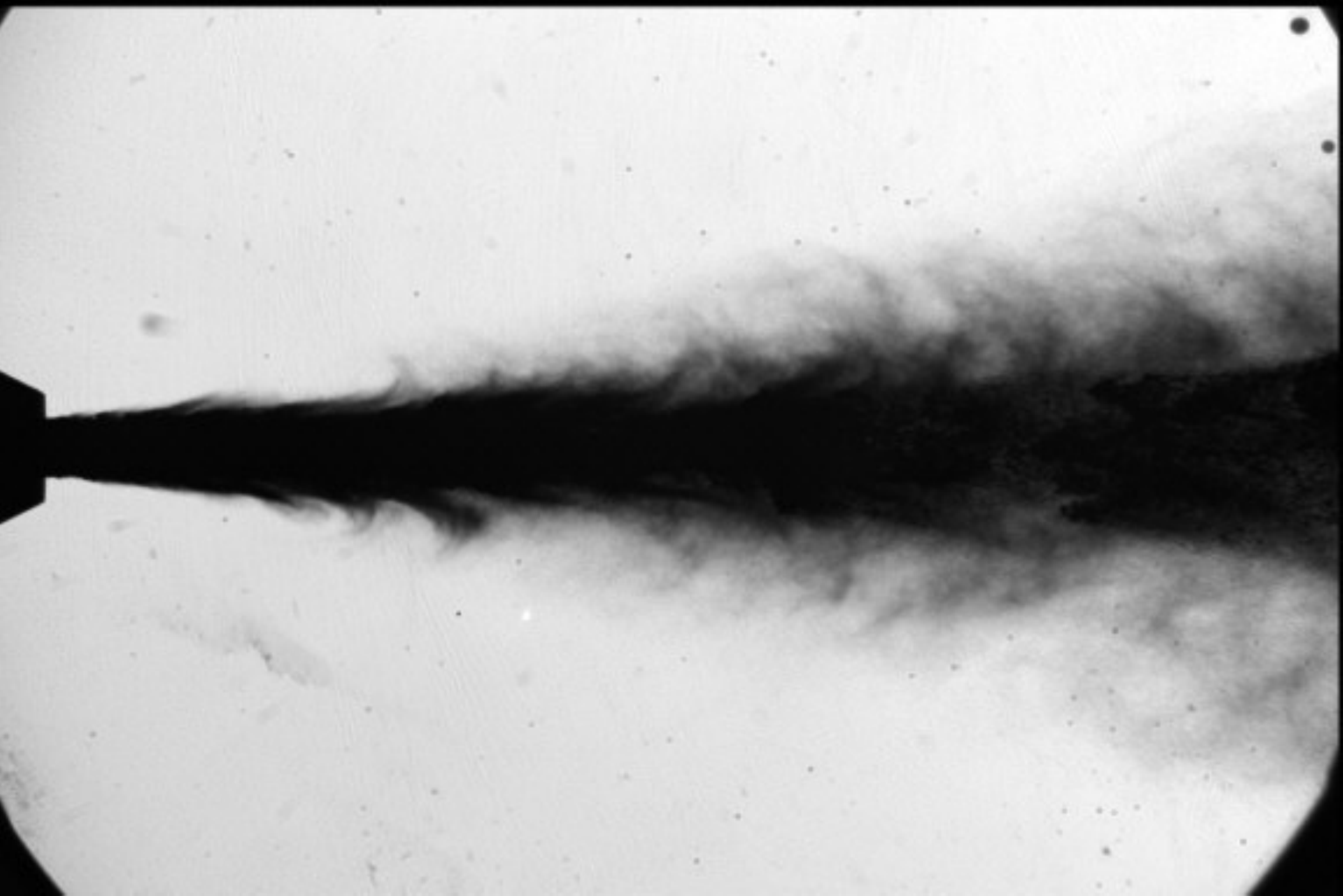}
\end {center}
\caption
{
Coaxial liquid-air jet; the
credit 
is due to   Mixing Enhancement via Secondary Parallel Injection  (MESPI).
}
\label{liquidairjet}
    \end{figure*}

Turbulent jets  are  a subject of laboratory 
experiments, as an example, see Figure \ref{liquidairjet}.
The theory of turbulent jets emerging from a
circular hole can be found in different books with different
theories, see \cite{foot}, \cite{landau}, and \cite{goldstein}.
The basic assumptions  common to the three
already cited approaches  are
\begin{enumerate}
\item  The rate of momentum flow, $J$,
represented by
\begin{equation}
J = constant \times \rho b_j^2 \overline {v}_{x,max}^2
\quad ,
\end{equation}
is constant; 
here $x$ is the distance from the initial circular hole, 
$b_j(x)$ is the jet's diameter at  distance $x$,
$\overline{v}_{x,max}$  is  the maximum velocity along the the
centreline, $constant$  is 
\begin{equation}
constant = 2 \pi  \int_0^{\infty} f^2 \xi d\xi 
\quad ,
\label{constant}
\end{equation}
where 
\begin{equation}
f(\xi) =  \frac {\overline {v}_x  }{\overline {v}_{x,max}  } 
 \quad with \quad \xi=\frac{x}{b_{1/2}}
\quad ,
\end{equation}
 and
$\rho$  is the density of the surrounding medium,  see equation
(5.6-3) in \cite{foot}. 
\item The jet's density $\rho$ is constant
      over the expansion
      and equal to that  of the surrounding medium.
      The pressure is absent in this theory.
\end{enumerate}

Omitting the  details of the computation,  
an expression can be found
for the average velocity  $\overline{v}_{x}$,
see equation (5.6-21) in \cite{foot},
\begin{equation}
\overline{v}_{x}  =
\frac{\nu^{(t)} }{x}
\frac {2 C_3^2}{ \bigl [ 1 +\frac {1}{4}(C_3 \frac {r}{x})^2 \bigr ]^2}
\quad ,
\end{equation}
where  $\nu^{(t)}$ is the kinematical
eddy viscosity and  $C_3$ is as follows, see equation (5.6-23) in \cite{foot},
\begin{equation}
C_3 =
{\sqrt { \frac {3}{16 \pi}}
{\sqrt { \frac {J} {\rho}}}}
{\frac {1} { \nu^{(t)}}}
\quad .
\label{astroviscosity}
\end{equation}
An important quantity is the radial position, $r=b_{1/2}$,
corresponding to an axial velocity one-half of the centreline
value, see equation (5.6-24) in \cite{foot},
\begin{equation}
\frac
{\overline {v}_x (b_{1/2},x) }
{\overline {v}_{x,max}  (x) }
= \frac{1}{2} =
\frac {1}{ \bigl [ 1 +\frac {1}{4}(C_3 \frac {b_{1/2}}{x})^2 \bigr ]^2}
\label{ratiovel}
\quad .
\end{equation}
The experiments in the range of Reynolds number, $Re$,
$10^4 \le Re \le 3 10^6 $ (see  
\cite{Reichardt1942},
\cite{Reichardt1951} and \cite{Schlichting1979})
indicate that
\begin{equation}
b_{1/2}= 0.0848 x
\quad ,
\label{bvalueturb}
\end{equation}
and as  a consequence
\begin{equation}
C_3 = 15.17
\quad ,
\label{cmvalue1}
\end{equation}
and therefore
\begin{equation}
\frac
{\overline {v}_x (r) }
{\overline {v}_{x,max}  (r) }  =
\frac {1}{ \bigl [ 1 + 0.414 (\frac {r}{b_{1/2}})^2 \bigr ]^2}
\label{ratiovelnumeri}
\quad .
\end{equation}
The average velocity, $\overline{v}_{x}$,
 is   $\approx ~ 1/100$ of the centreline
value when $r/b_{1/2}$ = 4.6  and this allows seeing that the
diameter  of the jet is
\begin{equation}
b_j =  2 \times 4.6 b_{1/2}
\quad .
\end{equation}
On  introducing the opening angle $\alpha$, 
the following new 
relation is found:
\begin{equation}
\frac { \alpha }{2} = \arctan \frac { 4.6 b_{1/2}}{x} \quad .
\label{newalfa}
\end{equation}
The generally accepted relation 
between  the opening angle and Mach number, see equation (A33) in
\cite{deyoung}, is 
\begin{equation}
\frac { \alpha }{2} = \frac{c_s}{v_j} = \frac {1}{M}  \quad ,
\label{oldalfa}
\end{equation}
where $c_s$ is the velocity of sound, $v_j$ the jet's velocity, 
and  $M$ the Mach number.
The new relation~(\ref{newalfa})  replaces 
the traditional relation~(\ref{oldalfa}). 
The parameter $b_{1/2}$  can therefore be connected
with the jet's geometry: 
\begin{equation}
b_{1/2}=
\frac {1}{4.6}
\tan (\frac { \alpha }{2}) x
\quad .
\end{equation}
If  this approximate theory  is accepted,
 equation~(\ref{bvalueturb}) gives
$\alpha=42.61^{\circ}$: this is the theoretical 
value that yields the so called
Reichardt profiles.
The value of  $b_{1/2}$ fixes the value of $C_3$ and therefore
the eddy viscosity is
\begin{equation}
\nu^{(t)} =
{\sqrt { \frac {3}{16 \pi}}
{\sqrt { \frac {J} {\rho}}}}
{\frac {1} { C_3 }}
=
{\sqrt { \frac {3}{16 \pi}}}
{\sqrt {constant}}
\;
{ b {v}_{x,max}}
{\frac {1}{C_3}}
\quad .
\end{equation}
In order to continue, the integral 
 that appears  in $constant$
 should be evaluated,
 see equation~(\ref{constant}).
Numerical integration gives 
\begin{equation}
\int_0^{\infty} f^2 \xi d\xi =0.402 
\quad ,
\end{equation}
and therefore
\begin{equation}
constant=2.528 
\quad .
\end{equation}
On introducing typical parameters of jets  like
$\alpha$=$5^{\circ}$,
${v}_{x,max}$=$v_{100}=v[{\mathrm{km/sec}}]/100$, 
$b_j=b_1$, where $b_1$ is the
momentary diameter in ${\mathrm{ pc}}$,
it 
is possible to deduce an astrophysical  formula for
the kinematical eddy viscosity:
\begin{equation}
\nu^{(t)} =
2.92\,10^{-9}\, b_1 v_{100}  \mathrm{ \frac {pc^2}{year}}
\quad  when
\quad C_3 = 135.61
\quad .
\end{equation}
This paragraph  concludes by underlining   the fact   that
in extra-galactic sources it is possible to observe 
both a small opening angle,  $\approx~5^{\circ}$
and 
great opening angles, i.e. $\approx~34^{\circ}$ in the outer 
regions of 3C31 \cite{LaingBridle2004}.

\subsection{The Lane--Emden profile}

The self gravitating sphere of a polytropic
gas is governed
by the Lane--Emden differential equation
of the second order
\begin{equation}
{\frac {d^{2}}{d{x}^{2}}}Y ( x ) +2\,{\frac {{\frac {d}{dx}
}Y ( x ) }{x}}+ ( Y ( x )  ) ^{n}=0
\quad ,
\nonumber
\end{equation}
where $n$  is an integer, see
\cite{Lane1870,Emden1907,Chandrasekhar_1967,Binney2011,Zwillinger1989}.
The solution  $Y ( x )_n$
has  the density profile
\begin{equation}
\rho = \rho_c Y ( x )_n^n
\quad ,
\nonumber
\end{equation}
where $\rho_c$ is the density at $x=0$.
The pressure $P$ and temperature $T$ scale as
\begin{equation}
P = K \rho^{1 +\frac{1}{n}}
\quad ,
\label{pressure}
\end{equation}
\begin{equation}
 T = K^{\prime} Y(x)
\label{temperature}
\quad ,
\end{equation}
where  K and K$^{\prime}$   are two  constants.
For more details, see \cite{Hansen1994}.

Analytical solutions exist for $n=0$, 1, and 5.
The analytical  solution for $n$=5 is
\begin{equation}
Y(x) ={\frac {1}{{(1+ \frac{{x}^{2}}{3})^{1/2}}} }
\quad ,
\nonumber
\end{equation}
and the density for $n$=5 is
\begin{equation}
\rho(x) =\rho_c {\frac {1}{{(1+ \frac{{x}^{2}}{3})^{5/2}}} }
\label{densitan5}
\quad .
\end{equation}

The   variable  $x$   is   non-dimensional
and  we now  introduce the
new variable $x=r/b$
\begin{equation}
\rho(r;b) =\rho_c {\frac {1}{{(1+ \frac{{r}^{2}}{3b^2})^{5/2}}} }
\label{densita5b}
\quad .
\end{equation}

\subsection{Preliminaries}

The chosen  physical units are
pc for length  and  yr for time;
with these units, the initial velocity $v_{{0}}$
is  expressed in pc yr$^{-1}$.
When the initial velocity is expressed in
km\,s$^{-1}$, the multiplicative factor $1.02\times10^{-6}$
should be applied in order to have the velocity expressed in
pc yr$^{-1}$.
In these units, the speed of light is
$c=0.306$  \ pc \ yr$^{-1}$.
The    goodness of the approximation of a solution is evaluated
through the percentage error, $\delta$, which is
\begin{equation}
\delta = \frac{\big | x - x_{app}|}
{x} \times 100
\quad ,
\end{equation}
where $x$ is the analytical or numerical solution 
and  $x_{app}$ the approximate solution, see \cite{Abramowitz1965}.

\subsection{Classical solution to first order}

\begin{figure*}
\begin{center}
\includegraphics[width=7cm]{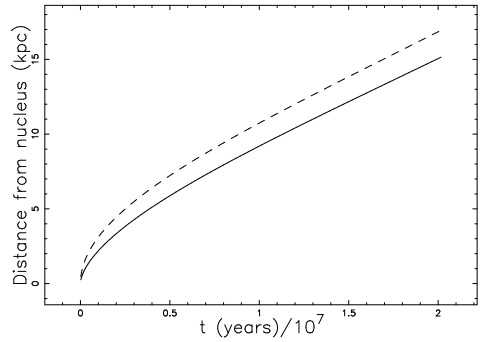}
\end {center}
\caption
{
Numerical solution  as given
by equation (\ref{xt}) (full line)
and asymptotic  solution
 as given
by equation (\ref{xtasymptotic}) (dashed line),
with parameters as in Table \ref{jetparameters}.
}
\label{traj_asymp}
    \end{figure*}

The conservation of the energy flux in a straight  turbulent jet
the concept of the perpendicular section
section to the motion along the
Cartesian $x$-axis, $A$
\begin {equation}
A(r)=\pi~r^2
\end{equation}
where $r$ is the radius of the jet.
The
section  $A$ at  position $x_0$  is
\begin {equation}
A(x_0)=\pi ( x_0   \tan ( \frac{\alpha}{2}))^2
\end{equation}
where   $\alpha$  is the opening angle and
$x_0$ is the initial position on the $x$-axis.
At position $x$ we have
\begin {equation}
A(x)=\pi ( x   \tan ( \frac{\alpha}{2}))^2
\label{ax}
\quad .
\end{equation}
The conservation  of energy flux states that
\begin{equation}
\frac{1}{2} \rho(x_0)  v_0^3   A(x_0)  =
\frac{1}{2} \rho(x  )   v(x)^3 A(x)
\label{conservazioneenergy}
\end {equation}
where $v(x)$ is the velocity at  position $x$ and
$v_0(x_0)$   is the velocity at  position $x_0$,
see Formula A28 in \cite{deyoung}.
We now  assume that 
a  Lane--Emden  ($n=5$) density profile is valid, see
equation (\ref{densita5b}). Then 
the above conservation law  becomes   

\begin{eqnarray}
\frac{1}{2}\,{\rho_{{0}}{v(x)}^{3}\pi\,{x}^{2} \left( \tan \left( \frac{\alpha}{2}
 \right)  \right) ^{2} \left( 1+\frac{1}{3}\,{\frac {{x}^{2}}{{b}^{2}}}
 \right) ^{-5/2}}
\nonumber \\
=\frac{1}{2}\,{\rho_{{0}}{v_{{0}}(x_0)}^{3}\pi\,{x_{{0}}}^{2}
 \left( \tan \left( \frac{\alpha}{2} \right)  \right) ^{2} \left( 1+\frac{1}{3}\,{
\frac {{x_{{0}}}^{2}}{{b}^{2}}} \right) ^{-5/2}}
\quad ,
\end{eqnarray}
where $v(x)$ is the velocity at  position $x$,
$v_0(x_0)$   is the velocity at  position $x_0$
and
$\alpha$  is the opening angle of the jet.
The above equation is a cubic equation 
which 
has  one real root plus 
two non-real complex conjugate roots.
Here and in the following we take  into account only the real root. 
The  real analytical  solution for the velocity without losses is
\begin{equation}
v(x;b,x_0,v_0) = 
\frac
{
v_{{0}} \left( 3\,{b}^{2}+{x}^{2} \right) ^{{\frac{5}{6}}}{x_{{0}}}^{{
\frac{2}{3}}}
}
{
 \left( 3\,{b}^{2}+{x_{{0}}}^{2} \right) ^{{\frac{5}{6}}}{x}^{{\frac{2
}{3}}}
}
\quad .
\label{vfirst}
\end{equation}
The asymptotic expansion of above velocity, $v_a$,  
with respect to the variable $x$, which means $x\rightarrow \infty$, 
is  
\begin{equation}
v_a(x;b,x_0,v_0)
=
\frac
{
v_{{0}}{x_{{0}}}^{{\frac{2}{3}}} \left( 5\,{b}^{2}+2\,{x}^{2} \right)
}
{
2\, \left( 3\,{b}^{2}+{x_{{0}}}^{2} \right) ^{5/6}x
}
\label{vfirstasymptotic}
\quad .
\end{equation}
The trajectory can be found by the indefinite 
integral of
the inverse of the velocity as  given by equation (\ref{vfirst}): 
\begin{equation}
F(x)=\int   \frac{1}{v(x;b,x_0,v_0)} dx 
=
\frac
{
\sqrt [6]{3} \left( 3\,{b}^{2}+{x_{{0}}}^{2} \right) ^{{\frac{5}{6}}}{
x}^{{\frac{5}{3}}}
{\mbox{$_2$F$_1$}({\frac{5}{6}},{\frac{5}{6}};\,{\frac{11}{6}};\,-{\frac
{{x}^{2}}{3\,{b}^{2}}})}
}
{
5\,v_{{0}} \left( {b}^{2} \right) ^{5/6}{x_{{0}}}^{2/3}
}
\quad ,
\end{equation} 
where ${\2F1(a,b;\,c;\,v)}$ is a regularized hypergeometric function,
see \cite{Abramowitz1965,Seggern1992,Thompson1997,NIST2010}.
The trajectory expressed in terms of  $t$ as a function  of $x$
is  
\begin{equation}
F(x) - F(x_0) = t
\quad  .
\label{xt}
\end{equation}
The  above equation can not be inverted in the usual form,
which is  $x$ as a function of $t$.
The asymptotic trajectory can be found by the indefinite 
integral of
the inverse of the asymptotic  velocity as  given by 
equation (\ref{vfirstasymptotic}) 
\begin{equation}
F_a(x)=\int   \frac{1}{v_a(x;b,x_0,v_0)} dx 
=
\frac
{
 \left( 3\,{b}^{2}+{x_{{0}}}^{2} \right) ^{{\frac{5}{6}}}\ln  \left( 5
\,{b}^{2}+2\,{x}^{2} \right) 
}
{
2\,v_{{0}}{x_{{0}}}^{2/3}
}
\quad .
\end{equation} 
The equation of the asymptotic trajectory is
\begin{equation}
F_a(x) - F_a(x_0) = t
\quad  ,
\end{equation}
and the solution for $x$ of the above equation, 
the asymptotic trajectory, is
\begin{equation}
x(t;b,x_0,v_0)=
\frac{1}{2}
\sqrt {-10\,{b}^{2}+2\,{{\rm e}^{{\frac { \left( 3\,{b}^{2}+{x_{{0}}}^
{2} \right) ^{5/6}\ln  \left( 5\,{b}^{2}+2\,{x_{{0}}}^{2} \right) +2\,
tv_{{0}}{x_{{0}}}^{2/3}}{ \left( 3\,{b}^{2}+{x_{{0}}}^{2} \right) ^{5/
6}}}}}}
\quad .
\label{xtasymptotic}
\end{equation} 
Figure  \ref{traj_asymp} shows a typical example
of the above asymptotic  expansion.

\begin{table}[ht!]
\caption
{
Parameters for a classical extra-galactic jet.
}
\label{jetparameters}
\begin{center}
\begin{tabular}{|c|c|}
\hline
parameter    &  value    \\
\hline
$x_0$ (pc)   & 100       \\
$v_0$ ($\frac{km}{s}$)   & 10000       \\
$b$   (pc)   & 10000\\
\hline
\end{tabular}
\end{center}
\end{table}

\subsection{Solution to second order}

Let us suppose that the radiative losses
are  proportional to the   flux of energy 
\begin{equation}
- \epsilon 
\frac
{
\rho_{{0}}{v}^{3}\pi\,{x}^{2} \left( \tan \left( {\frac {\alpha}{2}}
 \right)  \right) ^{2}
}
{
2\, \left( 1+\frac{1}{3}\,{\frac {{x}^{2}}{{b}^{2}}} \right) ^{5/2}
}
\quad .
\end{equation}

\begin{figure*}
\begin{center}
\includegraphics[width=7cm]{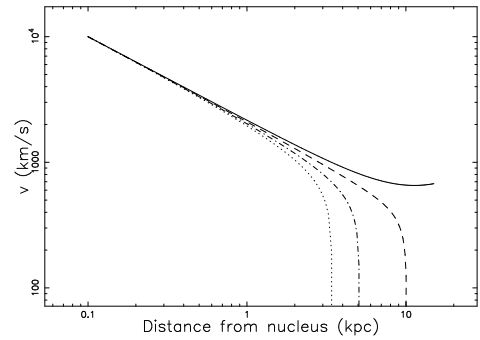}
\end {center}
\caption
{
Velocity corrected for radiative losses, i.e. velocity to second order, 
equation (\ref{vcorrected}), as a function of the distance,
with parameters as in Table \ref{jetparameters}:
$\epsilon=0 $ full   line,
$\epsilon=1.0\,10^{-4}$ dashed line,
$\epsilon=2.0\,10^{-4}$ dot-dash-dot-dash line
and  
$\epsilon=3.0\,10^{-4}$ dotted line.
}
\label{vel_emden_back}
    \end{figure*}

Inserting in the above equation  the  velocity to first order
as  given by equation~(\ref{vfirst}) 
the radiative losses, $Q(x;x_0,v_0,b,\epsilon)$, are
\begin{equation}
Q(x;x_0,v_0,b,\epsilon)= - \epsilon 
\frac
{
\rho_{{0}}{v}^{3}\pi\,{x}^{2} \left( \tan \left( {\frac {\alpha}{2}}
 \right)  \right) ^{2}
}
{
2\, \left( 1+\frac{1}{3}\,{\frac {{x}^{2}}{{b}^{2}}} \right) ^{5/2}
}
\quad ,
\label{lossesclassical}
\end{equation}
where $\epsilon$ is a constant which  fixes   the conversion 
of  the   flux of energy  to other kinds of energies, in this
case, the radiative losses.
The sum of the radiative  losses between $x_0$ and $x$ 
is given by the following integral, $L$,
\begin{equation}
L(x;x_0,v_0,b,\epsilon)=\int_{x_0}^x  Q(x;x_0,v_0,b,\epsilon) dx
=\frac
{
-9\,\epsilon\,\rho_{{0}}\sqrt {3}{b}^{5}{v_{{0}}}^{3}{x_{{0}}}^{2}\pi
\, \left( \tan \left( \alpha/2 \right)  \right) ^{2} \left( x-x_{{0}}
 \right) 
}
{
2\, \left( 3\,{b}^{2}+{x_{{0}}}^{2} \right) ^{5/2}
}
\quad .
\label{classiclosses}
\end{equation}
The  conservation of the   flux of energy  in the presence  
of  the back-reaction due to the radiative losses
is 
\begin{eqnarray}
\frac
{
9\,\sqrt {3}\rho_{{0}} \left( {b}^{5}{v_{{0}}}^{3}{x_{{0}}}^{2}
\epsilon\, \left( {\frac {3\,{b}^{2}+{x}^{2}}{{b}^{2}}} \right) ^{5/2}
x-{b}^{5}{v_{{0}}}^{3}{x_{{0}}}^{3}\epsilon\, \left( {\frac {3\,{b}^{2
}+{x}^{2}}{{b}^{2}}} \right) ^{5/2}+{v}^{3}{x}^{2} \left( 3\,{b}^{2}+{
x_{{0}}}^{2} \right) ^{5/2} \right) 
}
{
2\, \left( {\frac {3\,{b}^{2}+{x}^{2}}{{b}^{2}}} \right) ^{5/2}
 \left( 3\,{b}^{2}+{x_{{0}}}^{2} \right) ^{5/2}
}
\nonumber \\
=
{
9\,\rho_{{0}}\sqrt {3}{v_{{0}}}^{3}{x_{{0}}}^{2}
}
{
2\, \left( {\frac {3\,{b}^{2}+{x_{{0}}}^{2}}{{b}^{2}}} \right) ^{5/2}
}
\quad  .
\label{consfluxback}
\end{eqnarray}
The  analytical solution for the velocity to 
second order, $v_c(x;b,x_0,v_0)$, 
is
\begin{equation}
v_c(x;b,x_0,v_0)=
\frac
{
v_{{0}}\sqrt [3]{1+\epsilon\, \left( -x+x_{{0}} \right) } \left( 3\,{b
}^{2}+{x}^{2} \right) ^{{\frac{5}{6}}}{x_{{0}}}^{{\frac{2}{3}}}
}
{
\left( 3\,{b}^{2}+{x_{{0}}}^{2} \right) ^{{\frac{5}{6}}}{x}^{{\frac{2
}{3}}}
}
\label{vcorrected}
\quad ,
\end{equation}
and  Figure \ref{vel_emden_back}  shows 
an example.

\begin{figure*}
\begin{center}
\includegraphics[width=7cm]{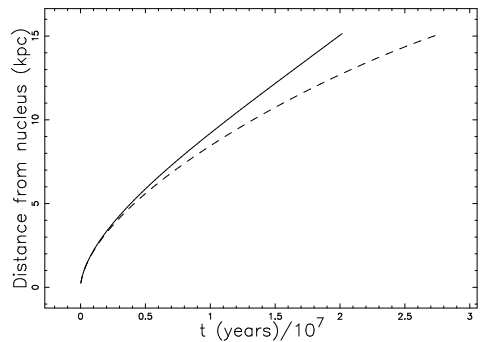}
\end {center}
\caption
{
Numerical trajectory  corrected for radiative losses 
as a function of time,
with parameters as in Table \ref{jetparameters}:
$\epsilon=0 $ full   line and
$\epsilon=8.0\,10^{-5}$ dashed line.
}
\label{traj_back}
    \end{figure*}

There are no analytical  results for the 
trajectory corrected for radiative losses,
and a numerical example is shown
in Figure \ref{traj_back}.

\begin{figure*}
\begin{center}
\includegraphics[width=7cm]{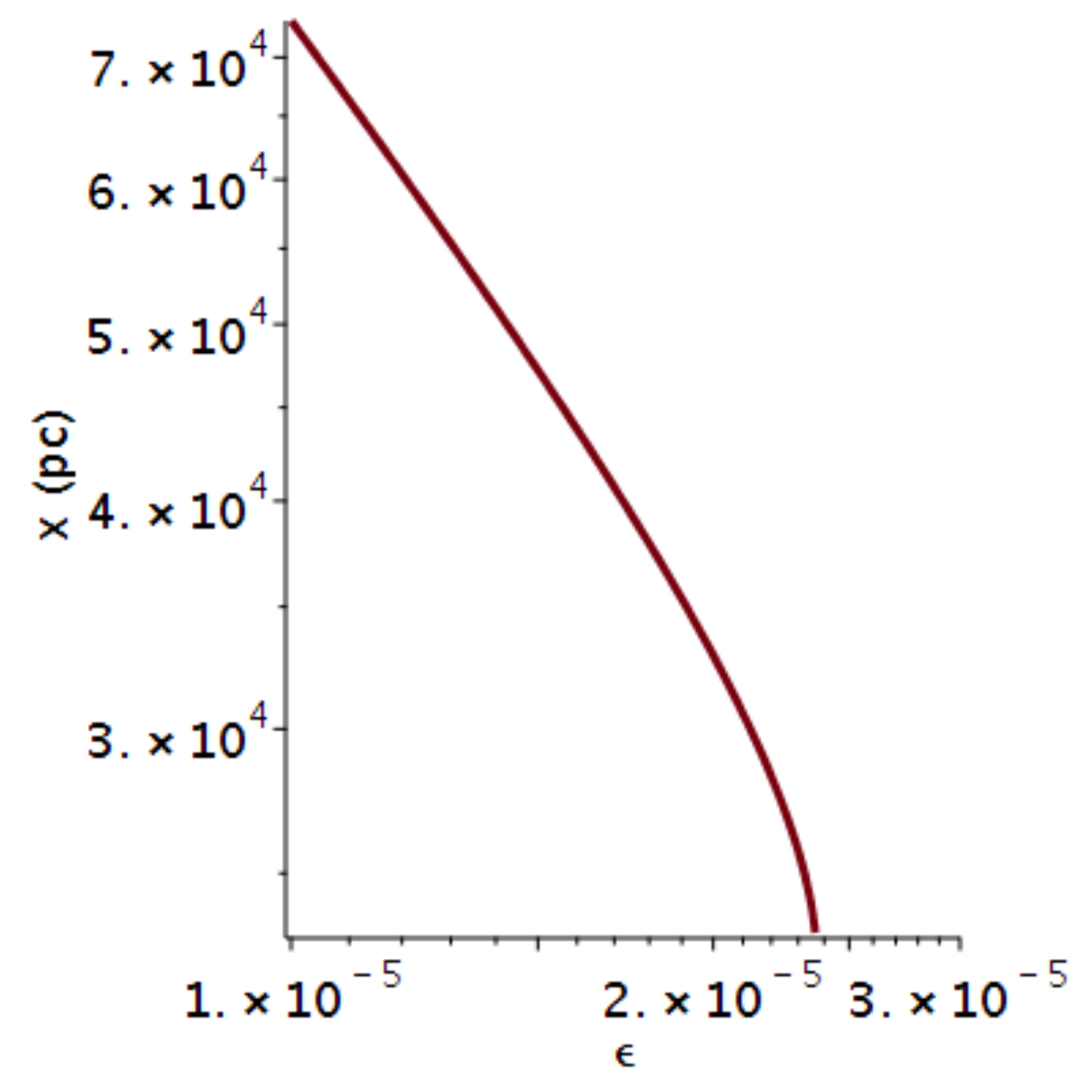}
\end {center}
\caption
{
Length  of the  jet, $x_j$,  in pc   as a function of $\epsilon$,
with $b$ as  in Table \ref{jetparameters}.
}
\label{xlunpc}
    \end{figure*}

The  inclusion  of back-reaction  allows the evaluation of the 
jet's length,  which can be derived from the minimum  
in the corrected velocity to second order as a function of $x$, 
\begin{equation}
\frac{\partial v_c(x;b,x_0,v_0)}{\partial x} =0
\quad ,
\end {equation}

which is  
\begin{eqnarray}
-{\frac {v_{{0}}\epsilon}{3} \left( 3\,{b}^{2}+{x}^{2} \right) ^{{
\frac{5}{6}}}{x_{{0}}}^{{\frac{2}{3}}} \left( 1+\epsilon\, \left( -x+x
_{{0}} \right)  \right) ^{-{\frac{2}{3}}} \left( 3\,{b}^{2}+{x_{{0}}}^
{2} \right) ^{-{\frac{5}{6}}}{x}^{-{\frac{2}{3}}}}
\nonumber \\
+{\frac {5\,v_{{0}}
}{3}\sqrt [3]{1+\epsilon\, \left( -x+x_{{0}} \right) }{x_{{0}}}^{{
\frac{2}{3}}}\sqrt [3]{x} \left( 3\,{b}^{2}+{x_{{0}}}^{2} \right) ^{-{
\frac{5}{6}}}{\frac {1}{\sqrt [6]{3\,{b}^{2}+{x}^{2}}}}}
\nonumber \\
-{\frac {2\,v_
{{0}}}{3}\sqrt [3]{1+\epsilon\, \left( -x+x_{{0}} \right) } \left( 3\,
{b}^{2}+{x}^{2} \right) ^{{\frac{5}{6}}}{x_{{0}}}^{{\frac{2}{3}}}
 \left( 3\,{b}^{2}+{x_{{0}}}^{2} \right) ^{-{\frac{5}{6}}}{x}^{-{\frac
{5}{3}}}}=0
\quad  .
\end{eqnarray}
The solution for $x$ of the above minimum determines
the jet's length, $x_j$,
\begin{equation}
x_j =
\frac
{
4\,{b}^{2}{\epsilon}^{2}+{\epsilon}^{2}{x_{{0}}}^{2}+\sqrt [3]{D_{{2}}
}\epsilon\,x_{{0}}+{D_{{2}}}^{{\frac{2}{3}}}+2\,\epsilon\,x_{{0}}+
\sqrt [3]{D_{{2}}}+1
}
{
4\,\epsilon\,\sqrt [3]{D_{{2}}}
}
\quad ,
\end{equation}
where
\begin{eqnarray}
D_1=
-16\,{b}^{4}{\epsilon}^{4}+429\,{b}^{2}{\epsilon}^{4}{x_{{0}}}^{2}-24
\,{\epsilon}^{4}{x_{{0}}}^{4}+858\,{b}^{2}{\epsilon}^{3}x_{{0}}
\nonumber\\
-96\,{
\epsilon}^{3}{x_{{0}}}^{3}
+429\,{b}^{2}{\epsilon}^{2}-144\,{\epsilon}^
{2}{x_{{0}}}^{2}-96\,\epsilon\,x_{{0}}-24
\quad ,
\end{eqnarray}  
and
\begin{equation}
D_2=
-42\,{b}^{2}{\epsilon}^{3}x_{{0}}+{\epsilon}^{3}{x_{{0}}}^{3}-42\,{b}^
{2}{\epsilon}^{2}
+3\,{\epsilon}^{2}{x_{{0}}}^{2}+2\,b\sqrt {{\it D1}}
\epsilon+3\,\epsilon\,x_{{0}}+1
\quad  .
\end{equation}
Figure \ref{xlunpc} shows $x_j$  numerically.

\section{Conservation of relativistic flux of energy}

\label{secrelativistic}

The corrections in special relativity (SR)
for stable atomic clocks in satellites
of the
Global Positioning System (GPS) are applied to 
satellites which are moving
at a velocity of $\approx 3.87 \frac{km}{s} $, 
see \cite{Ashby2003,Ashby2009}.

In astrophysics we deal with velocities near that of light
and therefore we should introduce   relativistic conservation 
laws.
The conservation of the     relativistic   flux of energy 
in  SR
in the   presence of a  velocity $v$ along one direction
states that
\begin{equation}
A(x) \frac { 1}{ 1 -\frac {v^2}{c^2}} (e_0 +p_0) v = cost
\label{enthalpy}
\end{equation}
where $A(x)$ is the considered area in the direction perpendicular
to the motion,
$c$ is the speed of light,
$e_0= c^2 \rho$ is the energy density in the rest 
frame of the moving fluid,
and $p_0$ is the pressure in the rest frame 
of the moving fluid,
see  formula A31 in
\cite{deyoung} and  \cite{Zaninetti2016e}.
In accordance with the current models of classical turbulent jets,
we  insert $p_0=0$.  Then
the   conservation law
for   the   relativistic   flux of energy
is

\begin{equation}
\rho  c^2 v \frac { 1}{ 1 -\frac {v^2}{c^2} } A(x) = cost
\quad .
\label{relativisticflux}
\end{equation}

\begin{figure*}
\begin{center}
\includegraphics[width=7cm]{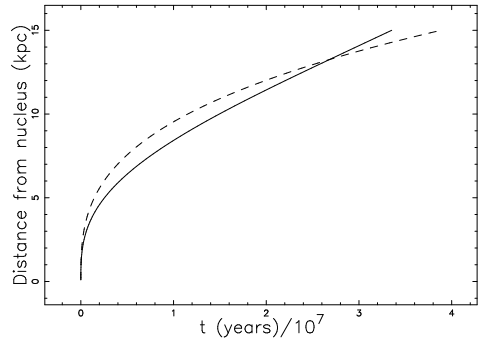}
\end {center}
\caption
{
Numerical relativistic solution  as given
by equation (\ref{eqnmotionrel}) (full line)
and truncated series expansion
as given
by equation (\ref{traj_rel_series}) (dashed line),
with parameters as in Table \ref{jetrel}.
}
\label{traj_rel_series}
    \end{figure*}

In the presence of a 
Lane--Emden  ($n=5$) density profile, as given
by equation (\ref{densita5b}) and $A(x)$ as given
by equation~(\ref{ax}),
the conservation  
of     relativistic   flux of energy for 
a straight jet  
takes the form 
\begin{equation}
\frac
{
\rho_{{0}}{c}^{3}\beta\,\pi\,{x}^{2} \left( \tan \left( \alpha/2
 \right)  \right) ^{2}
}
{
\left( 1+\frac{1}{3}\,{\frac {{x}^{2}}{{b}^{2}}} \right) ^{5/2} \left(1- {\beta
}^{2} \right) 
}
=
\frac
{
\rho_{{0}}{c}^{3}\beta0\,\pi\,{x_{{0}}}^{2} \left( \tan \left( \alpha
/2 \right)  \right) ^{2}
}
{
 \left( 1+\frac{1}{3}\,{\frac {{x_{{0}}}^{2}}{{b}^{2}}} \right) ^{5/2} \left( 
{1-\beta0}^{2} \right) 
}
\quad ,
\end{equation}
where  
$v$ is the velocity at $x$,
$v_0$ is the velocity at $x_0$,
$\beta=\frac{v}{c}$ and $\beta_0=\frac{v_0}{c}$.
The solution for $\beta$ to first order is 
\begin{equation}
\beta(x;x_0,b,\beta_0)=\frac
{N}
{
 \left( 1+\frac{1}{3}\,{\frac {{x_{{0}}}^{2}}{{b}^{2}}} \right) ^{5/2} \left( 
{\beta0}^{2}-1 \right) 
}
\quad  ,
\label{betarelfirst}
\end{equation}
where
\begin{eqnarray}
N=
9\,\sqrt {3\,{b}^{2}+{x_{{0}}}^{2}}{x}^{2}{b}^{4}{\beta_{{0}}}^{2}+6\,
\sqrt {3\,{b}^{2}+{x_{{0}}}^{2}}{x}^{2}{b}^{2}{x_{{0}}}^{2}{\beta_{{0}
}}^{2}+\sqrt {3\,{b}^{2}+{x_{{0}}}^{2}}{x}^{2}{x_{{0}}}^{4}{\beta_{{0}
}}^{2}
\nonumber \\
-9\,\sqrt {3\,{b}^{2}+{x_{{0}}}^{2}}{x}^{2}{b}^{4}-6\,\sqrt {3\,
{b}^{2}+{x_{{0}}}^{2}}{x}^{2}{b}^{2}{x_{{0}}}^{2}-\sqrt {3\,{b}^{2}+{x
_{{0}}}^{2}}{x}^{2}{x_{{0}}}^{4}
\nonumber \\
+\Biggl (243\,{x}^{4}   ( {b}^{2}+ \frac{1}{3}
\,{x_{{0}}}^{2}   ) ^{5}{\beta_{{0}}}^{4}+   ( -2\,{x}^{4}{x_{
{0}}}^{10}-30\,{b}^{2}{x}^{4}{x_{{0}}}^{8}-180\,{b}^{4}{x}^{4}{x_{{0}}
}^{6}
\nonumber \\
+   ( 972\,{b}^{10}+1620\,{b}^{8}{x}^{2}+540\,{b}^{6}{x}^{4}+
360\,{b}^{4}{x}^{6}+60\,{b}^{2}{x}^{8}+4\,{x}^{10}   ) {x_{{0}}}^{
4}
\nonumber \\
-810\,{b}^{8}{x}^{4}{x_{{0}}}^{2}-486\,{b}^{10}{x}^{4}   ) {
\beta_{{0}}}^{2}+243\,{x}^{4}   ( {b}^{2}+\frac{1}{3}\,{x_{{0}}}^{2}
   ) ^{5}\Biggr )^{1/2}
\quad .
\end{eqnarray}
The equation for the relativistic trajectory   is
\begin{equation}
\int_{x_0}^x \frac{1}{\beta(x;x_0,b,\beta_0) \, c} dx  = t
\quad .
\label{eqnmotionrel}
\end{equation} 
The integral in the above equation does not have an analytical
solution 
and should be integrated numerically.
In order to have analytical results, 
two approximation are now introduced.
The first approximation  computes a truncated series expansion
for the integrand of the integral in equation (\ref{eqnmotionrel}),
which transforms the relativistic equation of motion
into 
\begin{equation}
F(x) - F(x_0) = t
\label{eqn_traj_rel_series}
\quad ,
\end{equation}
with  
\begin{equation}
F(x) = \frac{NF}{162\,{x_{{0}}}^{2}\beta_{{0}}{b}^{10}c}
\quad ,
\end{equation}
where 
\begin{eqnarray}
NF =  \left( {b}^{2} \right) ^{5/2}x \Bigl( 9\,\sqrt {3}\sqrt {3\,{b}^{2}+
{x_{{0}}}^{2}}{b}^{4}{\beta_{{0}}}^{2}{x}^{2}+6\,\sqrt {3}\sqrt {3\,{b
}^{2}+{x_{{0}}}^{2}}{b}^{2}{\beta_{{0}}}^{2}{x}^{2}{x_{{0}}}^{2}
\nonumber \\
+
\sqrt {3}\sqrt {3\,{b}^{2}+{x_{{0}}}^{2}}{\beta_{{0}}}^{2}{x}^{2}{x_{{0
}}}^{4}-9\,\sqrt {3}\sqrt {3\,{b}^{2}+{x_{{0}}}^{2}}{b}^{4}{x}^{2}
\nonumber  \\
-6\,
\sqrt {3}\sqrt {3\,{b}^{2}+{x_{{0}}}^{2}}{b}^{2}{x}^{2}{x_{{0}}}^{2}-
\sqrt {3}\sqrt {3\,{b}^{2}+{x_{{0}}}^{2}}{x}^{2}{x_{{0}}}^{4}-162\,
\sqrt {{b}^{10}{x_{{0}}}^{4}{\beta_{{0}}}^{2}} \Bigr ) 
\quad  .
\end{eqnarray}
In the above analytical  result we have the time as 
a function of the distance, 
see  Figure~\ref{traj_rel_series} 
where the percentage error at $x=15$\ kpc is
$\delta= 15.91\%$.

\begin{table}[ht!]
\caption
{
Parameters for a relativistic extra-galactic jet.
}
\label{jetrel}
\begin{center}
\begin{tabular}{|c|c|}
\hline
parameter    &  value    \\
\hline
$x_0$ (pc)   & 100       \\
$\beta_0$    & 0.9       \\
$b$   (pc)   & 10000\\
\hline
\end{tabular}
\end{center}
\end{table}

\begin{figure*}
\begin{center}
\includegraphics[width=7cm]{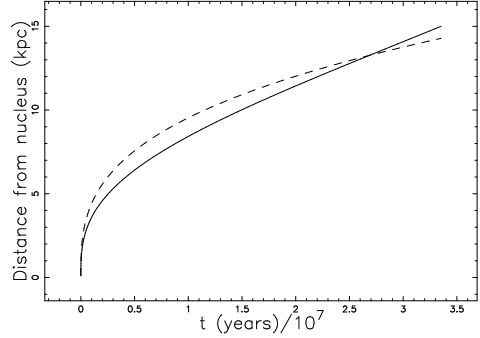}
\end {center}
\caption
{
Numerical relativistic solution  as given
by equation (\ref{eqnmotionrel}) (full line)
and Pad\'e approximant 
 as given
by equation (\ref{traj_rel_pade}) (dashed line),
with parameters as in Table \ref{jetrel}.
}
\label{traj_rel_pade}
    \end{figure*}

The second approximation  computes a 
Pad\'e  approximant of order  [2/1], 
see \cite{2012Adachi,Aviles2014,Wei2014}, 
for the integrand of the integral in equation (\ref{eqnmotionrel})
\begin{equation}
P(x) - P(x_0) = t
\quad ,
\end{equation}
with  
\begin{equation}
P(x) \frac{NP}{162\,{b}^{10}{x_{{0}}}^{4}{\beta_{{0}}}^{2}c}
\quad ,
\end{equation}
where
\begin{eqnarray}
NP(x) = -x \left( {b}^{2} \right) ^{5/2}{x_{{0}}}^{2}\beta_{{0}} \Big ( 9\,
\sqrt {3}\sqrt {3\,{b}^{2}+{x_{{0}}}^{2}}{b}^{4}{\beta_{{0}}}^{2}{x}^{
2}
\nonumber  \\
+6\,\sqrt {3}\sqrt {3\,{b}^{2}+{x_{{0}}}^{2}}{b}^{2}{\beta_{{0}}}^{2
}{x}^{2}{x_{{0}}}^{2}+\sqrt {3}\sqrt {3\,{b}^{2}+{x_{{0}}}^{2}}{\beta_
{{0}}}^{2}{x}^{2}{x_{{0}}}^{4}
\nonumber \\
-9\,\sqrt {3}\sqrt {3\,{b}^{2}+{x_{{0}}}
^{2}}{b}^{4}{x}^{2}-6\,\sqrt {3}\sqrt {3\,{b}^{2}+{x_{{0}}}^{2}}{b}^{2
}{x}^{2}{x_{{0}}}^{2}-\sqrt {3}\sqrt {3\,{b}^{2}+{x_{{0}}}^{2}}{x}^{2}
{x_{{0}}}^{4}
\nonumber \\
-162\,\sqrt {{b}^{10}{x_{{0}}}^{4}{\beta_{{0}}}^{2}}
 \Big ) 
\quad .
\end{eqnarray}
The above equation can be inverted, but the 
analytical expression for $x=G(t;x_0,\beta_0,b)$ as a function of time 
is complicated
and is omitted here.
As an example, with the parameters of Table \ref{jetrel},
we have 
\begin{equation}
G(t) =\frac{NG}{DG}
\label{eqn_traj_rel_pade}
\quad ,
\end{equation}
with 
\begin{eqnarray}
NG =
-{ 2.9237\times 10^{-17}}\, \Bigl ( -{ 1.7397\times 10^{54}}
\,t-{ 5.8851\times 10^{56}}
\nonumber \\
+{ 2.9816\times 10^{20}}\,\sqrt {
{ 1.9201\times 10^{73}}+{ 2.3032\times 10^{70}}\,t
}
\nonumber  \\
{
+{
 3.4042\times 10^{67}}\,{t}^{2}} \Bigr ) ^{2/3}+{ 3.2399
\times 10^{21}}
\quad ,
\end{eqnarray}
and 
\begin{eqnarray}
DG=
\Bigg (
-{ 1.7397\times 10^{54}}\,t-{ 5.8851\times 10^{56}
}
\nonumber \\
+{ 2.9816\times 10^{20}}\,\sqrt {{ 1.9201\times 10^{73}}+{
 2.3032\times 10^{70}}\,t+{ 3.4042\times 10^{67}}\,{t}^{2}}
\Bigg )^{\frac{1}{3}}
\quad .
\end{eqnarray}
An example is shown in 
Figure \ref{traj_rel_pade}, where the percentage error at $x=15$\ kpc is
$\delta= 4.81\%$.

\subsection{Relativistic solution to second order}

\begin{figure*}
\begin{center}
\includegraphics[width=7cm]{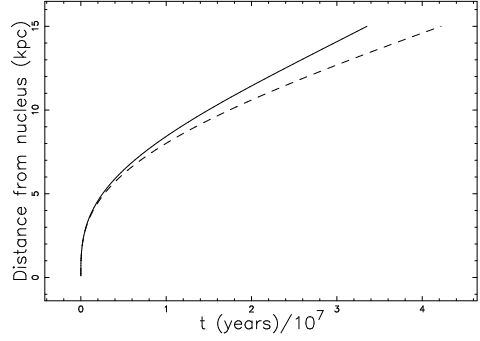}
\end {center}
\caption
{
Numerical relativistic solution  as given
by equation (\ref{eqnmotionrel}) (full line)
and solution with back-reaction, i.e. to second order,  
 (dashed line),
with parameters as in Table \ref{jetrel}
and $\epsilon=2.0\,10^{-5}$.
}
\label{traj_back_eps}
    \end{figure*}

We now  suppose that the radiative losses
are  proportional to the    relativistic   flux of energy.
The integral of the losses, $L_r$, between $x_0$ and $x$ is
\begin{equation}
L_r(x;x_0,\beta_0,b,c) = - \epsilon 
\frac
{
9\, \left( x-x_{{0}} \right) \rho0\,{c}^{3}\beta_{{0}}{x_{{0}}}^{2}
\pi\, \left( \tan \left( \alpha/2 \right)  \right) ^{2}{b}^{5}\sqrt {3
}
}
{
\left( 3\,{b}^{2}+{x_{{0}}}^{2} \right) ^{5/2} \left(1- {\beta_{{0}}}^{
2} \right) 
}
\quad .
\label{relativisticlosses}
\end{equation}
The  conservation of the    relativistic   flux of energy in the presence  
of  the back-reaction due to the radiative losses
is 
\begin{eqnarray}
\frac{NR}
{
\left( 3\,{b}^{2}+{x}^{2} \right) ^{5/2} \left(
3\,{b}^{2}+{x_{{0}}}^
{2} \right) ^{5/2} \left( {\beta}^{2}-1 \right)  \left( {\beta_{{0}}}^
{2}-1 \right) 
}=
\nonumber \\
\frac
{
9\,\rho0\,\sqrt {3}{c}^{3}\beta_{{0}}{x_{{0}}}^{2}{b}^{5}
}
{
 \left( 3\,{b}^{2}+{x_{{0}}}^{2} \right) ^{5/2} \left( {\beta_{{0}}}^{
2}-1 \right) 
}
\quad ,
\end{eqnarray}
where 
\begin{eqnarray}
NR = 81\,\rho0\,{b}^{5}\sqrt {3} \Bigg  (    ( {b}^{2}+\frac{1}{3}\,{x}^{2}
   ) ^{2}\epsilon\,   ( \beta+1   ) \beta_{{0}}   ( 
\beta-1   )    ( x-x_{{0}}   ) {x_{{0}}}^{2}\sqrt {3\,{b}^{
2}+{x}^{2}}
\nonumber \\
+   ( {b}^{2}+\frac{1}{3}\,{x_{{0}}}^{2}   ) ^{2}\beta\,{x}^
{2}   ( \beta_{{0}}+1   )    ( \beta_{{0}}-1   ) \sqrt {
3\,{b}^{2}+{x_{{0}}}^{2}} \Bigg  ) {c}^{3}
\quad .
\end{eqnarray}
The solution of the above equation, to second order, for $\beta$ 
is  
\begin{equation}
\beta= \frac{NB}
{
2\, \left( 3\,{b}^{2}+{x}^{2} \right) ^{5/2} \left(
\epsilon\,x-
\epsilon\,x_{{0}}-1 \right)  \left( 3\,{b}^{2}+{x_{{0}}}^{2} \right) {
x_{{0}}}^{2}\beta_{{0}}
}
\quad , 
\label{betarelsecond}
\end{equation}
where
\begin{eqnarray}
NB=
-\sqrt {3\,{b}^{2}+{x_{{0}}}^{2}}  \Bigg ( \sqrt {3\,{b}^{2}+{x_{{0}}}^{
2}} \times
\nonumber \\
\Big ( {{x}^{4}   ( \beta_{{0}}-1   ) ^{2}   ( \beta_{{0}}
+1   ) ^{2}{x_{{0}}}^{10}+15\,{b}^{2}{x}^{4}   ( \beta_{{0}}-1
   ) ^{2}   ( \beta_{{0}}+1   ) ^{2}{x_{{0}}}^{8}
}
\nonumber \\
{
+   ( 
972\,   ( {b}^{2}+\frac{1}{3}\,{x}^{2}   ) ^{5}{\beta_{{0}}}^{2}{
\epsilon}^{2}+90\,{b}^{4}{x}^{4}   ( \beta_{{0}}-1   ) ^{2}
   ( \beta_{{0}}+1   ) ^{2}   ) {x_{{0}}}^{6}
}
\nonumber \\
{
-1944\,
\epsilon\,   ( {b}^{2}+\frac{1}{3}\,{x}^{2}   ) ^{5}{\beta_{{0}}}^{2}
   ( \epsilon\,x-1   ) {x_{{0}}}^{5}+   ( 972\,   ( {b}^{
2}+\frac{1}{3}\,{x}^{2}   ) ^{5}{\beta_{{0}}}^{2}{x}^{2}{\epsilon}^{2}
}
\nonumber \\
{
-
1944\,   ( {b}^{2}+\frac{1}{3}\,{x}^{2}   ) ^{5}{\beta_{{0}}}^{2}x
\epsilon+4\,{x}^{10}{\beta_{{0}}}^{2}+60\,{b}^{2}{x}^{8}{\beta_{{0}}}^
{2}+360\,{b}^{4}{x}^{6}{\beta_{{0}}}^{2}
}
\nonumber \\
{
+270\,{b}^{6}   ( {\beta_{{0
}}}^{2}+1   ) ^{2}{x}^{4}+1620\,{b}^{8}{x}^{2}{\beta_{{0}}}^{2}+
972\,{b}^{10}{\beta_{{0}}}^{2}   ) {x_{{0}}}^{4}
}
\nonumber \\
{
+405\,{b}^{8}{x}^{
4}   ( \beta_{{0}}-1   ) ^{2}   ( \beta_{{0}}+1   ) ^{2}
{x_{{0}}}^{2}+243\,{b}^{10}{x}^{4}   ( \beta_{{0}}-1   ) ^{2}
   ( \beta_{{0}}+1   ) ^{2}}
\Big )^{1/2}
\nonumber \\
+27\,   ( {b}^{2}+\frac{1}{3}\,{x_{{0}}}
^{2}   ) ^{3}   ( \beta_{{0}}+1   ) {x}^{2}   ( \beta_{{0
}}-1   )   \Bigg ) 
\quad .
\end{eqnarray}
The relativistic equation of motion with back-reaction 
can be solved by numerically integrating the  relation
in equation (\ref{eqnmotionrel}).  
Figure \ref{traj_back_eps} gives an example.

\section{Astrophysical applications}

\label{secapplication}
We now analyse two models for the synchrotron emission
along the jet.

\subsection{Direct conversion}

The flux of observed radiation  
along the centre of the jet, $I_c$,  
in the classical case is  assumed to scale   
as
\begin{equation} 
I_c(x;x_0,v_0,b,\epsilon)\propto \frac{L(x;x_0,v_0,b,\epsilon)}{x^2}
\label{classicintensity}
\quad  ,
\end{equation}
where  $L$,
the sum of the radiative  losses, 
 is given by equation (\ref{classiclosses}).

The above relation  connects  the observed 
intensity of radiation with the rate of energy transfer per unit area.
In the relativistic case
\begin{equation} 
I_c(x;x_0,\beta_0,b,c)\propto \frac{L_r(x;x_0,\beta_0,b,c)}{x^2}
\quad  ,
\label{relativisticintensity}
\end{equation}
where  $L_r$ is given by equation (\ref{relativisticlosses})

A statistical test for the 
the goodness of fit
is the 
observational percentage  of
reliability,
 $\epsilon_{\mathrm {obs}}$,
\begin{equation}
\epsilon_{\mathrm {obs}}  =100\bigl (1-\frac{ \sum_j |I_{obs}-I_{theo}|_j }
                                      { \sum_j I_{theo,j}           }
                              \bigr )
.
\label{efficiencymany}
\end{equation}

\begin{figure*}
\begin{center}
\includegraphics[width=7cm]{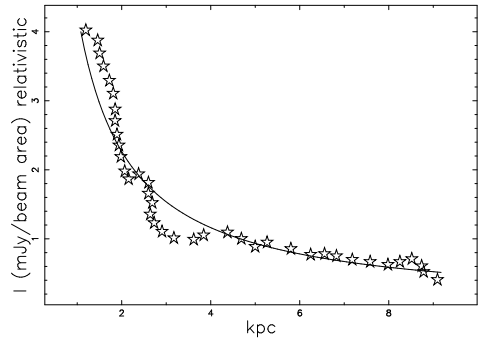}
\end {center}
\caption
{
Observed intensity profile along the centreline
of 3C31  (empty stars) 
and  theoretical intensity as given
by equation (\ref{relativisticintensity}), 
with  parameters as in Table \ref{jetrel} (full line).
The observational percentage  of
reliability is
$\epsilon_{\mathrm {obs}} =  86.19\%$.
}
\label{intensityrel_3c31}
    \end{figure*}

In order to make a comparison with the observed profile 
of intensity, we chose  3C31,  
see Figure 8 in
\cite{laing2002};
Figure \ref{intensityrel_3c31} 
shows the theoretical synchrotron
intensity as well as the observed one.

\subsection{The magnetic field of equipartition}

\begin{figure*}
\begin{center}
\includegraphics[width=7cm]{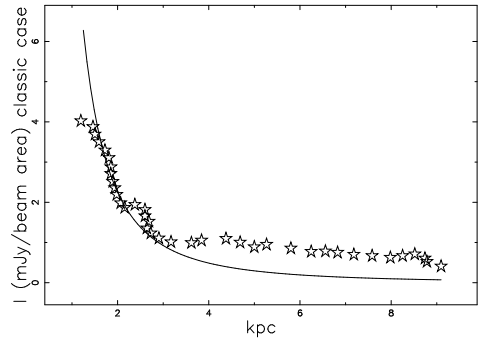}
\end {center}
\caption
{
Observed intensity profile along the centreline, $I_c$,
of 3C31  (empty stars) 
and  theoretical intensity as given
by equation 
(\ref{intensitycenterline}), with parameters 
as in Table \ref{jetparameters}.
The observational percentage  of
reliability is
$\epsilon_{\mathrm {obs}} =  73.79\%$.
}
\label{intensitsynchro_3c31}
    \end{figure*}

The magnetic field  in CGS has an energy
density of $\frac{B^2}{8 \pi}$
where $B$ is the magnetic field.
The presence of the magnetic field can be modeled 
assuming equipartition between  the kinetic energy 
and the magnetic energy
\begin{equation}
\frac{B(x)^2}{8 \pi} = \frac{1}{2} \rho v^2
\quad .
\end{equation}
Inserting the above equation 
in the classical equation for the conservation
of the flux of energy
 (\ref{conservazioneenergy}), a factor 2
will appear on both sides of the equation, leaving
unchanged the result for the deduction of the velocity to first order. 
The magnetic field
as a function of the distance $x$ 
when  the velocity is  given by equation (\ref{vfirst})
and in the presence of a Lane--Emden ($n=5$) profile for the density
is
\begin{equation}
B(x;x_0,b) =
\frac
{
B_{{0}} \left( 3\,{b}^{2}+{x_{{0}}}^{2} \right) ^{{\frac{5}{12}}}{x_{{0
}}}^{{\frac{2}{3}}}
}
{
\left( 3\,{b}^{2}+{x}^{2} \right) ^{{\frac{5}{12}}}{x}^{{\frac{2}{3}}
}
}
\label{bx}
\quad .
\end{equation}
where $B_0$ is the magnetic field at $x=x_0$.
We  assume   an inverse power law  spectrum
for the ultrarelativistic  electrons,
of the type
\begin{equation}
N(E)dE = K E^{-p} dE
\label{spectrum}
\end{equation}
where $K$ is a constant and $p$ the exponent of the inverse power law.
The intensity of the synchrotron radiation has a standard
expression, as given
by formula (1.175)  in \cite{lang2},
\begin{eqnarray}
I(\nu)
\approx 0.933 \times 10^{-23}
\alpha_p (p) K l  H_{\perp} ^{(p +1)/2 }
\bigl (
 \frac{6.26 \times 10^{18} }{\nu}
\bigr )^{(p-1)/2 }  \\
erg\, sec^{-1} cm^{-2} Hz^{-1} rad^{-2}
\nonumber
\label{isynchro}
\end{eqnarray}
where $\nu$ is the frequency,
$H_{\perp}$ is the magnetic field perpendicular to the 
electron's velocity,
$l$ is the dimension of the radiating region
along the line of sight,
and  $ \alpha_p (p)$  is a slowly
varying function
of $p$ which is of the order of unity.
We now analyse the intensity along the centreline of the jet,
which means that the  radiating length is
\begin{equation}
l(x;\alpha) = x   \tan ( \frac{\alpha}{2})
\quad .
\end{equation}
The intensity, assuming a constant $p$, scales as
\begin{equation}
I(x;x_0,p)=
\frac
{
I_{{0}}{B}^{{\frac {p}{2}}+{\frac{1}{2}}}x
}
{
{B_{{0}}}^{{\frac {p}{2}}+{\frac{1}{2}}}{\it x_0}
}
\quad , 
\end{equation}
where $I_0$  is the intensity at  $x=x_0$ and  $B_0$ 
the magnetic field
at $x=x_0$.
We insert Eq.~(\ref{bx})  
in order to have an analytical
expression for the centreline intensity
\begin{equation}
I(x;x_0,p,b)=
 \left( 3\,{b}^{2}+{x_{{0}}}^{2} \right) ^{{\frac {5\,p}{24}}+{\frac{5
}{24}}}i_{{0}}{x}^{-{\frac {p}{3}}+{\frac{2}{3}}}{x_{{0}}}^{{\frac {p
}{3}}-{\frac{2}{3}}} \left( 3\,{b}^{2}+{x}^{2} \right) ^{-{\frac {5\,p
}{24}}-{\frac{5}{24}}}
\label{intensitycenter}
\quad .
\end{equation}
The above equation for the intensity is relative to 
the unit area; 
in order to have the intensity on the centreline,
$I_c$,
we should make a further division by the area 
of interest, which scales $\propto \,x^2$
\begin{equation}
I_c(x;x_0,p,b)= \frac{I(x;x_0,p,b)}{x^2}
\label{intensitycenterline}
\quad .
\end{equation}
Figure \ref{intensitsynchro_3c31} 
shows the theoretical synchrotron
intensity with the variable magnetic field 
as well as the observed one for 3C31.

\section{Conclusions}

{\bf Classical case} The approximate trajectory  
of a turbulent jet  
in the presence of a  Lane--Emden ($n=5$) medium 
has been evaluated to first order, see equation (\ref{xtasymptotic}).
The  solution for the velocity to first order 
allows the  insertion of the back-reaction, i.e.
the radiative losses, in the equation for the   flux of energy
conservation, 
see equation (\ref{consfluxback}),
and as a consequence  the  velocity corrected to second order,
see equation (\ref{vcorrected}).
The  trajectory, calculated numerically  to second order, is shown 
in Figure  \ref{traj_back}.
The radiative losses allow evaluating the length  at which 
the advancing velocity of the jet is zero.
This length  has a complicated analytical expression
and was presented numerically, see Figure \ref{xlunpc}.

\noindent
{\bf Relativistic case}
In the relativistic case it is  possible to derive 
an analytical expression for $\beta$ to first order, see equation 
(\ref{betarelfirst}), and
to second order (taking into account radiative losses),
see equation 
(\ref{betarelsecond}).
The relativistic  trajectory to first order has been  evaluated 
through  a series, see equation (\ref{eqn_traj_rel_series})
or a Pad\'e  approximant of order [2/1], 
see equation (\ref{eqn_traj_rel_pade}).
The relativistic equation of motion to second order (back-reaction) 
has been  evaluated numerically,  see   
Figure \ref{traj_back_eps}.
In other words, with the introduction of the radiative losses,
the length of the classical or relativistic 
jet becomes finite rather than infinite.

\noindent 
{\bf An astrophysical application}
The radiative losses are represented by 
equation (\ref{lossesclassical})  in the classical case 
and by (\ref{relativisticlosses}) in the relativistic case.
A division of the  two above quantities by 
the area 
of interest allows deriving 
the theoretical rate of energy transfer per unit area,
which can be compared with the intensity 
of radiation along the jet, for example, 3C31,
see Figure \ref{intensityrel_3c31}.
The  spatial behaviour of the magnetic field  is introduced under the hypothesis 
of equipartition between the kinetic and magnetic energy,
see equation~(\ref{bx}), and this  allows closing the
standard equation for the synchrotron emissivity, see equation
(\ref{isynchro}).


\begin{thebibliography}{-------}
\providecommand{\natexlab}[1]{#1}

\bibitem[{Curtis}(1918)]{Curtis1918}
{Curtis}, H.D.
\newblock {Descriptions of 762 Nebulae and Clusters Photographed with the
  Crossley Reflector}.
\newblock {\em Publications of Lick Observatory} {\bf 1918}, {\em 13},~9--42.

\bibitem[{Fanaroff} and {Riley}(1974)]{fanaroff}
{Fanaroff}, B.L.; {Riley}, J.M.
\newblock {The morphology of extragalactic radio sources of high and low
  luminosity}.
\newblock {\em \mnras} {\bf 1974}, {\em 167},~31P--36P.

\bibitem[{Kembhavi} and {Narlikar}(1999)]{Kembhavi1999}
{Kembhavi}, A.K.; {Narlikar}, J.V.
\newblock {\em {Quasars and active galactic nuclei : an introduction}};
  Cambridge University Press: Cambridge,  1999.

\bibitem[{Bridle} and {Perley}(1984)]{Bridle1984}
{Bridle}, A.H.; {Perley}, R.A.
\newblock {Extragalactic Radio Jets}.
\newblock {\em \araa} {\bf 1984}, {\em 22},~319--358.

\bibitem[{Liu} and {Zhang}(2002)]{Liu2002}
{Liu}, F.K.; {Zhang}, Y.H.
\newblock {A new list of extra-galactic radio jets}.
\newblock {\em \aap} {\bf 2002}, {\em 381},~757--760,
  \href{http://xxx.lanl.gov/abs/astro-ph/0212477}{{\normalfont
  [astro-ph/0212477]}}.

\bibitem[{Hardcastle} and {Sakelliou}(2004)]{Hardcastle2004}
{Hardcastle}, M.J.; {Sakelliou}, I.
\newblock {Jet termination in wide-angle tail radio sources}.
\newblock {\em \mnras} {\bf 2004}, {\em 349},~560--575.

\bibitem[{Laing} and {Bridle}(2002)]{laing2002}
{Laing}, R.A.; {Bridle}, A.H.
\newblock {Relativistic models and the jet velocity field in the radio galaxy
  3C 31}.
\newblock {\em \mnras} {\bf 2002}, {\em 336},~328--352.

\bibitem[{Nawaz} \em{et~al.}(2014){Nawaz}, {Wagner}, {Bicknell}, {Sutherland},
  and {McNamara}]{Nawaz2014}
{Nawaz}, M.A.; {Wagner}, A.Y.; {Bicknell}, G.V.; {Sutherland}, R.S.;
  {McNamara}, B.R.
\newblock {Jet-intracluster medium interaction in Hydra A - I. Estimates of jet
  velocity from inner knots}.
\newblock {\em \mnras} {\bf 2014}, {\em 444},~1600--1614,
  \href{http://xxx.lanl.gov/abs/1408.4512}{{\normalfont [1408.4512]}}.

\bibitem[{Nawaz} \em{et~al.}(2016){Nawaz}, {Bicknell}, {Wagner}, {Sutherland},
  and {McNamara}]{Nawaz2016}
{Nawaz}, M.A.; {Bicknell}, G.V.; {Wagner}, A.Y.; {Sutherland}, R.S.;
  {McNamara}, B.R.
\newblock {Jet-intracluster medium interaction in Hydra A - II. The effect of
  jet precession}.
\newblock {\em \mnras} {\bf 2016}, {\em 458},~802--815,
  \href{http://xxx.lanl.gov/abs/1602.02969}{{\normalfont [1602.02969]}}.

\bibitem[{Zaninetti}(2015)]{Zaninetti2015d}
{Zaninetti}, L.
\newblock {Classical and relativistic conservation of momentum flux in
  radio-galaxies }.
\newblock {\em Applied Physics Research} {\bf 2015}, {\em 7},~43--62.

\bibitem[{Zaninetti}(2016)]{Zaninetti2016e}
{Zaninetti}, L.
\newblock Classical and Relativistic Flux of Energy Conservation in
  Astrophysical Jets.
\newblock {\em Journal of High Energy Physics, Gravitation and Cosmology} {\bf
  2016}, {\em 1},~41--56.

\bibitem[{Kellermann} and {Richards}(1998)]{Kellermann1998}
{Kellermann}, K.I.; {Richards}, E.A.
\newblock {Radio Observations of the Hubble Deep Field}.
\newblock  The Hubble Deep Field; {Livio}, M.; {Fall}, S.M.; {Madau}, P., Eds.;
  Cambridge University Press: Cambridge,  1998; p.~60.

\bibitem[{Bird} \em{et~al.}(2002){Bird}, {Stewart}, and {Lightfoot}]{foot}
{Bird}, R.; {Stewart}, W.; {Lightfoot}, E.
\newblock {\em Transport Phenomena ; second Edition}; John Wiley and Sons: New
  York,  2002.

\bibitem[{Landau}(1987)]{landau}
{Landau}, L.
\newblock {\em Fluid Mechanics 2nd edition}; Pergamon Press: London,  1987.

\bibitem[{Goldstein}(1965)]{goldstein}
{Goldstein}, S.
\newblock {\em Modern Developments in Fluid Dynamics}; Dover: New York,  1965.

\bibitem[{Reichardt}(1942)]{Reichardt1942}
{Reichardt}, V.
\newblock {Gesetzmabigkeiten der freien Turbulenz}.
\newblock {\em VDI-Forschungsheft} {\bf 1942}, {\em 414},~141.

\bibitem[{Reichardt}(1951)]{Reichardt1951}
{Reichardt}, V.
\newblock {Vollstandige Darstellung der turbulenten Geschwindigkeitsverteilung
  in glatten Leitungen}.
\newblock {\em Zeitschrift fur Angewandte Mathematik und Mechanik} {\bf 1951},
  {\em 31},~208--219.

\bibitem[{Schlichting}(1979)]{Schlichting1979}
{Schlichting}, H.
\newblock {\em Boundary Layer Theory}; McGraw-Hill: New York,  1979.

\bibitem[{De Young}(2002)]{deyoung}
{De Young}, D.S.
\newblock {\em {The physics of extragalactic radio sources}}; University of
  Chicago Press: Chicago,  2002.

\bibitem[{Laing} and {Bridle}(2004)]{LaingBridle2004}
{Laing}, R.A.; {Bridle}, A.H.
\newblock {Adiabatic relativistic models for the jets in the radio galaxy 3C
  31}.
\newblock {\em \mnras} {\bf 2004}, {\em 348},~1459--1472,
  \href{http://xxx.lanl.gov/abs/astro-ph/0311499}{{\normalfont
  [astro-ph/0311499]}}.

\bibitem[{Lane}(1870)]{Lane1870}
{Lane}, H.J.
\newblock On the Theoretical Temperature of the Sun, under the Hypothesis of a
  gaseous Mass maintaining its Volume by its internal Heat, and depending on
  the laws of gases as known to terrestrial Experiment.
\newblock {\em American Journal of Science} {\bf 1870}, {\em 148},~57--74.

\bibitem[{Emden}(1907)]{Emden1907}
{Emden}, R.
\newblock {\em Gaskugeln: anwendungen der mechanischen w{a}rmetheorie auf
  kosmologische und meteorologische probleme}; B. Teubner.: Berlin,  1907.

\bibitem[{Chandrasekhar}(1967)]{Chandrasekhar_1967}
{Chandrasekhar}, S.
\newblock {\em {An introduction to the study of stellar structure}}; {Dover}:
  New York,  1967.

\bibitem[{Binney} and {Tremaine}(2011)]{Binney2011}
{Binney}, J.; {Tremaine}, S.
\newblock {\em {Galactic dynamics, Second Edition}}; Princeton University
  Press: Princeton, NJ,  2011.

\bibitem[{Zwillinger}(1989)]{Zwillinger1989}
{Zwillinger}, D.
\newblock {\em {Handbook of differential equations}}; Academic Press: New York,
   1989.

\bibitem[{Hansen} and {Kawaler}(1994)]{Hansen1994}
{Hansen}, C.J.; {Kawaler}, S.D.
\newblock {\em {Stellar Interiors. Physical Principles, Structure, and
  Evolution.}}; Springer-Verlag: Berlin,  1994.

\bibitem[{Abramowitz} and {Stegun}(1965)]{Abramowitz1965}
{Abramowitz}, M.; {Stegun}, I.A.
\newblock {\em {Handbook of Mathematical Functions with Formulas, Graphs, and
  Mathematical Tables}}; Dover: New York,  1965.

\bibitem[{von Seggern}(1992)]{Seggern1992}
{von Seggern}, D.
\newblock {\em CRC Standard Curves and Surfaces}; CRC: New York,  1992.

\bibitem[{Thompson}(1997)]{Thompson1997}
{Thompson}, W.J.
\newblock {\em Atlas for computing mathematical functions}; Wiley-Interscience:
  New York,  1997.

\bibitem[Olver \em{et~al.}(2010)Olver, Lozier, Boisvert, and Clark]{NIST2010}
Olver, F.W.J.e.; Lozier, D.W.e.; Boisvert, R.F.e.; Clark, C.W.e.
\newblock {\em {NIST handbook of mathematical functions.}}; {Cambridge
  University Press. }: Cambridge,  2010.

\bibitem[{Ashby}(2003)]{Ashby2003}
{Ashby}, N.
\newblock {Relativity in the Global Positioning System}.
\newblock {\em Living Reviews in Relativity} {\bf 2003}, {\em 6},~1.

\bibitem[{Ashby} and {Nelson}(2009)]{Ashby2009}
{Ashby}, N.; {Nelson}, R.A.
\newblock {GPS, Relativity, and Extraterrestrial Navigation}.
\newblock  IAU Symposium \#261, American Astronomical Society,  2009, Vol. 261,
  p. 889.

\bibitem[{Adachi} and {Kasai}(2012)]{2012Adachi}
{Adachi}, M.; {Kasai}, M.
\newblock {An Analytical Approximation of the Luminosity Distance in Flat
  Cosmologies with a Cosmological Constant}.
\newblock {\em Progress of Theoretical Physics} {\bf 2012}, {\em
  127},~145--152,  \href{http://xxx.lanl.gov/abs/1111.6396}{{\normalfont
  [arXiv:astro-ph.CO/1111.6396]}}.

\bibitem[{Aviles} \em{et~al.}(2014){Aviles}, {Bravetti}, {Capozziello}, and
  {Luongo}]{Aviles2014}
{Aviles}, A.; {Bravetti}, A.; {Capozziello}, S.; {Luongo}, O.
\newblock {Precision cosmology with Pad{\'e} rational approximations:
  Theoretical predictions versus observational limits}.
\newblock {\em \prd} {\bf 2014}, {\em 90},~043531,
  \href{http://xxx.lanl.gov/abs/1405.6935}{{\normalfont
  [arXiv:gr-qc/1405.6935]}}.

\bibitem[{Wei} \em{et~al.}(2014){Wei}, {Yan}, and {Zhou}]{Wei2014}
{Wei}, H.; {Yan}, X.P.; {Zhou}, Y.N.
\newblock {Cosmological applications of Pade approximant}.
\newblock {\em \jcap} {\bf 2014}, {\em 1},~45,
  \href{http://xxx.lanl.gov/abs/1312.1117}{{\normalfont
  [arXiv:astro-ph.CO/1312.1117]}}.

\bibitem[{Lang}(1980)]{lang2}
{Lang}, K.R.
\newblock {\em {Astrophysical formulae. (Second Edition)}}; Springer: New York,
   1980.

\end{thebibliography}

\end{document}